\documentclass[prd,showpacs]{revtex4}
\usepackage{graphicx}

\usepackage{amsmath,amsfonts,amssymb}

\def\be{\begin{equation}}
\def\ee{\end{equation}}
\def\bea{\begin{eqnarray}}
\def\eea{\end{eqnarray}}

\def\d{\rm{d}}

\def\S{\rm{S}}
\def\vr{\varrho}
\newcommand{\stg}{{\sqrt{-\tilde{g}}}}
\newcommand{\sg} {{\sqrt{-g}}}

\newcommand{\bi} {\begin{itemize}}
\newcommand{\ei} {\end{itemize}}
\newcommand{\ben} {\begin{enumerate}}
\newcommand{\een} {\end{enumerate}}

\newcommand{\del} {\partial}

\newcommand{\rar} {{\rightarrow}}

\newcommand{\la} {{\lambda}}
\newcommand{\al} {{\alpha}}
\newcommand{\ka} {{\kappa}}

\newcommand{\Om} {{\Omega}}
\newcommand{\om} {{\omega}}

\newcommand{\si} {{\sigma}}

\newcommand{\Ga} {{\Gamma}}
\newcommand{\ga} {{\gamma}}

\def\v1{\vspace{1cm}}
\def\be{\begin{equation}}
\def\ee{\end{equation}}
\def\bc{\begin{center}}
\def\ec{\end{center}}

\newcommand{\tidd} {{\ddot{\tilde{a}}}}
\newcommand{\tid} {{\dot{\tilde{a}}}}
\newcommand{\tia} {{\tilde{a}}}
\newcommand{\tit} {{\tilde{t}}}

\begin{document}

\title{Conformal relativity versus Brans-Dicke and superstring theories.}

\author{David Blaschke}
\email{blaschke@Physik.Uni-Bielefeld.DE}
\affiliation{Fakult\"at f\"ur Physik, Universit\"at Bielefeld,
Universit\"atstrasse, D-33615 Bielefeld, Germany}
\email{david.blaschke@physik.uni-rostock.de}
\affiliation{\it Fachbereich Physik, Universit\"at Rostock,
Universit\"atsplatz 3, D-18051 Rostock, Germany}

\author{Mariusz P. D\c{a}browski}
\email{mpdabfz@sus.univ.szczecin.pl}
\affiliation{\it Institute of Physics, University of Szczecin, Wielkopolska 15,
          70-451 Szczecin, Poland}

\date{\today}

\begin{abstract}

Conformal relativity theory which is also known as Hoyle-Narlikar theory
has recently been given some new interest. It is an extended relativity theory
which is invariant with respect to conformal transformations of the metric.

In this paper we show how conformal relativity is related to the Brans-Dicke theory
and to the low-energy-effective superstring theory. We show that conformal relativity action is
equaivalent to a transformed Brans-Dicke action for Brans-Dicke parameter $\omega =
-3/2$ in contrast to a reduced (graviton-dilaton) low-energy-effective superstring action
which corresponds to a Brans-Dicke action with Brans-Dicke parameter $\omega = -1$.
In fact, Brans-Dicke parameter $\omega = -3/2$ gives a border between
a standard scalar field and a ghost.

We also present basic cosmological solutions of conformal
relativity in both Einstein and string frames. The Eintein limit for flat
conformal cosmology solutions is unique and it is flat Minkowski space. This
requires the scalar field/mass evolution instead of the scale factor evolution
in order to explain cosmological redshift.

It is interesting that like in ekpyrotic/cyclic models, a possible transition through
a singularity in conformal cosmology in the string frame takes place in
the weak coupling regime.
\end{abstract}


\maketitle

\section{Introduction}
\setcounter{equation}{0}

It is well-known that some of the fundamental equations of physics
such as Maxwell equations or massless Dirac equation are invariant
with respect to conformal transformations of the metric
\cite{birell}. On the other hand, the Einstein equations and the massless
Klein-Gordon equation are not invariant with respect to these
transformations. However, a modification of these equations which
involves both metric and scalar field (scalar-tensor gravity) may
possess the property of conformal invariance. Such theories have
recently been given some new interest and called conformal
relativity theories. They are not exactly new since they were
studied already and their fundamental version is well-known under
the name of the Hoyle-Narlikar theory \cite{HN,chern,be74,narlikar,penrose}.
However, some new aspects have been added to them recently and, in particular,
geometrical evolution
of the universe was reinterpreted as an evolution of the
mass represented by a scalar field in a flat universe \cite{bbpp,iap,annalen}.
The idea is quite interesting and it can help to
resolve the problem of the dark energy in the universe
\cite{weinberg_kosmkonst,apj517,aj116,apj560,ptw}. Similar ideas
have been developed in yet another modification of general
relativity called Self Creation Cosmology \cite{barber} which also solves
the dark energy problem together with a series of other cosmological problems including
Pioneer spacecraft puzzle \cite{pioneer}.

In this paper we look for the connection between conformal
relativity and other alternative gravity theories such as
Brans-Dicke theory \cite{bd} and their overlap with unified
gauge-gravity theories superstring and M-theory
\cite{polchinsky,dab5,mar_string}. In fact, Brans-Dicke theory (with Dirac's
Large Numbers Hypothesis as a precursor \cite{dirac}), belongs to a
larger class of the scalar-tensor theories of gravity \cite{maeda}
whose properties has been recently studied in the context of
supernovae data and density perturbations \cite{scaltens,polarski}.

In Section \ref{conrel} we discuss the basic properties of
conformal invariance, present the field equations of conformal relativity and
show how they behave under conformal transformations of the metric.
In Section \ref{relation} we study the mutual relations between
conformal relativity, Brans-Dicke theory and low-energy-effective
superstring theories for a reduced graviton-dilaton spectrum. In
Section \ref{CC} we present the basic cosmological solutions of
conformal relativity for Friedmann universes in both the Einstein
and the string frames. In Section \ref{disc} we make a summary of
the results. More details of conformal relativity theory will be
studied in a separate paper \cite{annalen}.

A signature of the metric convention
(-+++) is applied throughout the paper, together with the Riemann and Ricci tensor convention
denoted +++ by Misner, Thorne and Wheeler \cite{mtw} (same conventions as in Hawking and Ellis \cite{hawk_ellis}).

\section{Conformal relativity}
\label{conrel}
\setcounter{equation}{0}

Suppose that we have two spacetime manifolds ${\cal M}, \tilde{\cal M}$ with
metrics $g_{\mu\nu}, \tilde{g}_{\mu\nu}$ and {\it the same} coordinates
$x^{\mu}$. We say that the two manifols are {\it conformal} to
each other if they are related by the following {\it conformal
transformation}
\bea
\label{conf_trafo}
\tilde{g}_{\mu\nu} &=& \Omega^2(x) g_{\mu\nu}~,
\eea
and the function $\Omega$ which is called a conformal factor must
be a twice-differentiable function of coordinates $x^{\mu}$ and
lie in the range $0<\Omega<\infty$.
The conformal transformations shrink or stretch the distances between the two points
described by the same coordinate system $x^{\mu}$ on the manifolds
${\cal M}, \tilde{\cal M}$, respectively, but they preserve the angles between vectors
(in particular null vectors which define light cones) which
leads to a conservation of the (global) causal structure of the manifold \cite{hawk_ellis}.
If we take $\Om =$ const. we deal with the so-called {\it scale
transformations} \cite{maeda}. In fact, conformal transformations are
{\it localized} scale transformations $\Om = \Om(x)$.

On the other hand, the {\it coordinate transformations} $x^{\mu} \to \tilde{x}^{\mu}$ only
relabel the coordinates and do not change geometry and
they are entirely {\it different} from conformal transformations \cite{narlikar}.
This is crucial since conformal
transformations lead to a {\it different physics} on
conformally related manifolds ${\cal M}, \tilde{\cal M}$ \cite{maeda}.
Since this will usually be related to a {\it different coupling}
of a physical field to gravity we will be talking about different
{\it frames} in which the physics is studied (see also Ref. \cite{flanagan} for
a slightly different view).

In $D=4$ spacetime dimensions the
determinant of the metric $g={\rm det}~g_{\mu\nu}$ transform as
\bea
\label{det}
\sqrt{-\tilde{g}} &=& \Omega^4 \sqrt{-g}~.
\eea
It is obvious from (\ref{conf_trafo}) that the following relations
for the inverse metrics and the spacetime intervals hold
\bea
\label{conf_trafo_inv}
\tilde{g}^{\mu\nu} &=& \Omega^{-2} g^{\mu\nu}~, \\
\d \tilde{s}^2 &=& \Omega^2 \d s^2~.
\eea
Finally, the notion of conformal flatness means that
\bea
\label{conf_flat}
\tilde{g}_{\mu\nu} \Omega^{-2}(x) &=& g_{\mu\nu} \equiv \eta_{\mu\nu}~,
\eea
where $\eta_{\mu\nu}$ is the flat Minkowski metric.

The application of (\ref{conf_trafo}) to the Christoffel connection coefficients
gives \cite{hawk_ellis}
\bea
\label{connections}
\tilde{\Ga}^{\la}_{\mu\nu}
&=&
\Ga^{\la}_{\mu\nu} + \frac{1}{\Om}\left( g^{\la}_{\mu} \Om_{,\nu} +
  g^{\la}_{\nu} \Om_{,\mu} - g_{\mu\nu}g^{\la\ka}\Om_{,\ka} \right)~,
\\
\label{connections1}
\Ga^{\la}_{\mu\nu}
&=&
\tilde{\Ga}^{\la}_{\mu\nu}- \frac{1}{\Om} \left( \tilde{g}^{\la}_{\mu}
  \Om_{,\nu} + \tilde{g}^{\la}_{\nu} \Om_{,\mu} -
  \tilde{g}_{\mu\nu}\tilde{g}^{\la\ka}\Om_{,\ka} \right)~.
\eea

The Ricci tensors and Ricci scalars in the two related frames $g_{\mu\nu}$ and
$\tilde{g}_{\mu\nu}$ transform as
\bea
\label{riccitensor1}
\tilde{R}_{\mu\nu}
&=&
R_{\mu\nu} + \Om^{-2}\left [
  4\Om_{,\mu}\Om_{,\nu}-\Om_{,\si}\Om^{,\si}g_{\mu\nu}\right ]
-\Om^{-1}\left [ 2\Om_{;\mu\nu}+\Box \Om g_{\mu\nu} \right ]~,
\\
\label{riccitensor2}
R_{\mu\nu} &=& \tilde{R}_{\mu\nu} - 3\Om^{-2} \Om_{,\rho}\Om^{,\rho}\tilde{g}_{\mu\nu}
+\Om^{-1}\left[2\Om_{;\mu\nu}+ \tilde{g}_{\mu\nu} \stackrel{\sim}{\Box} \Om  \right ]~,
\eea
\bea
\label{ricciscalar4}
\tilde{R} &=& \Om^{-2} \left [ R - 6\frac{\Box{\Om}}{\Om}
\right]~,\\
\label{ricciscalar5}
R &=& \Omega^2 \left[ \tilde{R} + 6
\frac{\stackrel{\sim}{\Box}\Om}{\Om} - 12 \tilde{g}^{\mu\nu}
\frac{\Om_{,\mu}\Om_{,\nu}}{\Om}\right ]~,
\eea
and the appropriate d'Alambertian operators change under (\ref{conf_trafo}) as
\bea
\stackrel{\sim}{\Box}\phi &=&\Om^{-2}\left( {\Box}\phi+
  2g^{\mu\nu}\frac{\Om_{,\mu}}{\Om}\phi_{,\nu} \right )~,\\
\Box\phi &=&\Om^{-2}\left( \stackrel{\sim}{\Box}\phi-
  2\tilde{g}^{\mu\nu}\frac{\Om_{,\mu}}{\Om}\phi_{,\nu} \right )~.
\eea
In these formulas the d'Alembertian $\stackrel{\sim}{\Box}$ taken with respect to the
metric $\tilde{g}_{\mu\nu}$ is different
from $\Box$ which is taken with respect to a conformally rescaled metric
$g_{\mu\nu}$.

An important feature of the conformal transformations is that they
preserve Weyl conformal curvature tensor
\bea
C_{\mu\nu\rho\sigma} &=& R_{\mu\nu\rho\sigma} + \frac{2}{D-2}
\left(g_{\mu[\sigma}R_{\rho]\nu} +
g_{\nu[\rho}R_{\sigma]\mu}\right) + \frac{2}{(D-1)(D-2)} R
g_{\mu[\rho}g_{\sigma]\mu}~,
\eea
which means that we have
\bea
\tilde{C}_{\mu\nu\rho\sigma} &=& C_{\mu\nu\rho\sigma}
\eea
under (\ref{conf_trafo}).

Let us remind that the vacuum Einstein-Hilbert action of general relativity reads as
\bea
\label{ehconf1}
{\S}_{EH} &=& \frac{1}{2\kappa^2}\int~\d^4x\stg \tilde{R}~,
\eea
where
\bea
\label{kappa4}
\kappa^2 &=& 8\pi G~.
\eea
Application of the conformal transformation (\ref{conf_trafo})
together with the help of (\ref{ricciscalar4}) yields (we have
assumed that $\kappa^2 = 6$ for a while)
\bea
\label{eh_conf2}
{\S}_{EH}
&=&
\frac{1}{2}\int~\d^4x\sg \Om^2\left ( \frac{1}{6} R-\frac{\Box\Om}{\Om}\right
)~
\eea
which means that the vacuum part of the Einstein-Hilbert
action is not invariant under conformal transformation (\ref{conf_trafo}), apart from
the case of the global transformations of the trivial type
$\tilde{g}_{\mu\nu} = {\rm const}. \times g_{\mu\nu}$.

However, a modification of the Einstein-Hilbert action
(\ref{ehconf1}) which allows for a scalar field $\tilde{\Phi}$ which reads as
\bea
\label{tildeconfinv}
\tilde{\S} &=&
 \frac{1}{2}\int~\d^4x\stg\tilde{\Phi}\left (
  \frac{1}{6}\tilde{R}\tilde{\Phi} -\stackrel{\sim}{\Box}\tilde{\Phi}\right )~,
\eea
together with the appropriate redefinition of the scalar field
\bea
\label{PhitildetoPhi}
\tilde{\Phi} &=& \Om^{-1} \Phi
\eea
is, in fact, conformally invariant since the conformally
transformed action has the same form, i.e.,
\bea
\label{confinv}
\S &=&
 \frac{1}{2}\int~\d^4x\sg{\Phi}\left (\frac{1}{6}{R}{\Phi}-{\Box}\Phi \right
 )~.
\eea
Now, one can see that the original form of the Einstein-Hilbert
action can be recovered from (\ref{tildeconfinv}) (or, alternatively (\ref{confinv})
provided we assume that
\bea
\label{kappaphi}
\kappa^2 &=& \frac{6}{\tilde{\Phi}^2} = \frac{6}{\varphi_0^2} =
{\rm const.}
\eea

The action (\ref{tildeconfinv}) (or similarly
(\ref{confinv})) is usually represented in a different form
by the application of the expression for a covariant d'Alambertian
for a scalar field in general relativity
\bea
\label{boxpartial}
\stackrel{\sim}{\Box}\tilde{\Phi} &=& \frac{1}{\sqrt{-\tilde{g}}}
\stackrel{\sim}{\partial}_{\mu} \left(\sqrt{-\tilde{g}}
\stackrel{\sim}{\partial}^{\mu}{\tilde{\Phi}}
\right)~,
\eea
which after integrating out the boundary term, gives \cite{maeda}
\bea
\label{boundarytilde}
\tilde{S} &=& \frac{1}{2} \int d^4x \sqrt{-\tilde{g}} \left[
\frac{1}{6} \tilde{R} \tilde{\Phi}^2 +
\stackrel{\sim}{\partial}_{\mu}\tilde{\Phi}\stackrel{\sim}{\partial}^{\mu}\tilde{\Phi}
\right]~,
\eea
and the second term is just a kinetic
term for the scalar field (cf. \cite{birell,hawk_ellis}). The
equations (\ref{boundarytilde}) are also conformally invariant since
the application of the formulas (\ref{det}), (\ref{ricciscalar4})
and (\ref{PhitildetoPhi}) together with the appropriate
integration of the boundary term gives the same form of the equations
\bea
\label{boundary}
S &=& \frac{1}{2} \int d^4x \sqrt{-g} \left[
\frac{1}{6} R \Phi^2 +
{\partial}_{\mu}\Phi {\partial}^{\mu}{\Phi}
\right]~.
\eea
Because of the type of non-minimal coupling of gravity to a scalar
field $\tilde{\Phi}$ or $\Phi$ in (\ref{boundarytilde}) and
(\ref{boundary}) and the relation to Brans-Dicke theory (see
Section (\ref{relation}) we say that these equations are presented in the {\it
Jordan frame} \cite{jordan,maeda}.

The conformally invariant actions (\ref{tildeconfinv}) and (\ref{confinv})
are the basis to derive the equations of motion via the variational principle.
The resulting equations of motion are conformally invariant, too.
The equations of motion for scalar fields $\tilde{\Phi}$ and $\Phi$
are conformally invariant
\bea
\label{eom_1}
\left
  (\stackrel{\sim}{\Box}-\frac{1}{6} \tilde{R} \right ) \tilde{\Phi} =
  \Om^{-3} \left(\Box - \frac{1}{6} R\right) \Phi &=& 0~,
\eea
and they have the structure of the Klein-Gordon equation with the
mass term replaced by the curvature term ~\cite{chern}.
The conformally invariant Einstein equations are obtained from
variation of $\tilde{S}$ with respect to the metric $\tilde{g}_{\mu\nu}$
and read as
\bea
\label{eom3}
\left( \tilde{R}_{\mu\nu}- \frac{1}{2}\tilde{g}_{\mu\nu}\tilde{R}
\right) \frac{1}{6} \tilde{\Phi}^2 + \frac{1}{6}
\left[ 4 \tilde{\Phi}_{,\mu}\tilde{\Phi}_{,\nu} -
\tilde{g}_{\mu\nu}
\tilde{\Phi}_{,\al}\tilde{\Phi}^{,\al} \right] + \frac{1}{3} \left[ \tilde{g}_{\mu\nu}
\tilde{\Phi} \stackrel{\sim}{\Box} \tilde{\Phi} -
\tilde{\Phi} \tilde{\Phi}_{\tilde{;}\mu\nu} \right] &=& 0~.
\eea

In order to prove the conformal invariance of the field equations (\ref{eom3})
it is necessary to know the rule of the conformal
transformations for the double covariant derivative of a scalar
field, i.e.,
\bea
\label{covdertilde}
\tilde{\Phi}_{\tilde{;}\mu\nu} &=& \tilde{\Phi}_{,\mu\nu} - \tilde{\Ga}^{\rho}_{\mu\nu}
\tilde{\Phi}_{,\rho} = - \Om^{-2} \Phi \Om_{;\mu\nu} + \Om^{-1}
\Phi_{;\mu\nu} + 4 \Om^{-3} \Phi \Om_{,\mu} \Om_{,\nu} \nonumber \\
&-& 2 \Om^{-2}
\left(\Phi_{,\mu}\Om_{,\nu} + \Om_{,\mu} \Phi_{,\nu} \right) -
\Om^{-3} \Phi g_{\mu\nu} \Om_{,\rho} \Om^{,\rho} + \Om^{-2}
g_{\mu\nu} \Phi_{,\rho} \Om^{,\rho} ~,
\eea
and
\bea
\label{covder}
\Phi_{;\mu\nu} &=& \tilde{\Phi} \Om_{;\mu\nu} + \Om \Phi_{;\mu\nu}
+ \frac{2}{\Om} \tilde{\Phi} \Om_{,\mu} \Om_{,\nu} +
2 \left( \Om_{,\mu} \tilde{\Phi}_{,\nu} + \tilde{\Phi}_{,\mu}
\Om_{,\nu} \right) - \frac{1}{\Om} \tilde{\Phi} \tilde{g}_{\mu\nu}
\Om_{\rho} \Om^{\rho} - \frac{1}{\Om} \tilde{g}_{\mu\nu}
\tilde{\Phi}_{,\rho} \Om^{,\rho} = 0~
\eea
($\tilde{;}$ means the covariant derivative with respect to
$\tilde{g}_{\mu\nu}$).

Inserting (\ref{riccitensor1}), (\ref{ricciscalar4}), (\ref{PhitildetoPhi})
and (\ref{covdertilde}) into
(\ref{eom3}) gives the same {\it conformally invariant} form of
the field equations as
\bea
\label{eom4}
\left( R_{\mu\nu}- \frac{1}{2}g_{\mu\nu}R
\right) \frac{1}{6} \Phi^2 + \frac{1}{6}
\left[ 4 \Phi_{,\mu}\Phi_{,\nu} -
g_{\mu\nu}
\Phi_{,\al} \Phi^{,\al} \right] + \frac{1}{3} \left[ g_{\mu\nu}
\Phi \Box \Phi -
\Phi \Phi_{;\mu\nu} \right] &=& 0~.
\eea
These are exactly the same field equations as in the Hoyle-Narlikar theory \cite{narlikar}.
Note that the scalar field equations of motion (\ref{eom_1}) can be obtained by the
appropriate contraction of equations (\ref{eom3}) and (\ref{eom4})
so that they are not independent and do not supply any additional
information \cite{canuto}.

Another point is that the equations (\ref{eom3}) or (\ref{eom4}) apparently could
give directly the vacuum Einstein field
equations for $\tilde{\Phi}= \varphi_0 = \sqrt{6}/\kappa = \sqrt{6/8\pi G} =$ const. (cf. Eq.
(\ref{kappaphi})). The same is obviously true for the field
equations (\ref{eom4}) with the same value of $\Phi = \varphi_0 = \sqrt{6}/\kappa =
\sqrt{6/8\pi G}$ = const. However, this limit is restricted to the
case of vanishing Ricci curvature $R=0$ or $\tilde{R}=0$ (so only flat Minkowski space limit is
allowed) which can
be seen from the scalar field equations of motion (\ref{eom_1}).

The admission of the matter part
\bea
\label{matteract}
\S_{\rm Matter} &=& \frac{1}{2} \int~\d^4x\sg~\cal{L}_{\rm Matter}~,
\eea
into the action (\ref{confinv}), with the matter energy-momentum tensor
\bea
\label{mattertrafo}
T^{\mu\nu} &=& \frac{2}{\sqrt{-g}}\frac{\del}{\del  g_{\mu\nu}}\left(
  \sqrt{-g}\cal{L}_{\rm Matter}\right )~,
\eea
allows to generalize the field equations (\ref{eom4}) to
\bea
\label{eom4T}
\left( R_{\mu\nu}- \frac{1}{2}g_{\mu\nu}R
\right) \frac{1}{6} \Phi^2 + \frac{1}{6}
\left[ 4 \Phi_{,\mu}\Phi_{,\nu} -
g_{\mu\nu}
\Phi_{,\al} \Phi^{,\al} \right] + \frac{1}{3} \left[ g_{\mu\nu}
\Phi \Box \Phi -
\Phi \Phi_{;\mu\nu} \right] &=&  T_{\mu\nu}~.
\eea
These equations (\ref{eom4T}), after contraction, give modified field equations (\ref{eom_1})
\bea
\label{eom_1T}
\left(\Box - \frac{1}{6} R\right) \Phi &=&
  \frac{T}{\Phi}~.
\eea
Note that putting $\Phi = \phi_0 = \sqrt{6/8\pi G}$ into
(\ref{eom4T}) gives Einstein field equations provided $T = -(1/6)R\Phi^2$.

In the conformal frame we add the matter term
\bea\label{emt}
\tilde{\S}_{\rm Matter} &=&  \frac{1}{2}\int~\d^4x~\stg~\Om^{-4}\tilde{\cal{L}}_{\rm
  Matter}~,
\eea
with
\bea
\label{mattertrafotilde}
\tilde{T}^{\mu\nu} &=&
\frac{2}{\sqrt{-\tilde{g}}}\frac{\partial}{\partial
  \tilde{g}_{\mu\nu}}\left( \sqrt{-\tilde{g}}\Om^{-4}\tilde{\cal{L}}_{\rm
    Matter}\right )~.
\eea
Some manipulations with the help of (\ref{conf_trafo}), (\ref{det}) and
(\ref{conf_trafo_inv}) show that
\bea
\tilde{T}^{\mu\nu} &=& \Omega^{-6} T^{\mu\nu}~,\\
\tilde{T} &=& \Om^{-4} T~.
\eea
The admission of (\ref{emt}) gives the conformally invariant field
equations (\ref{confinv}) as
\bea
\label{eom4Ttil}
\left( \tilde{R}_{\mu\nu}- \frac{1}{2}\tilde{g}_{\mu\nu}\tilde{R}
\right) \frac{1}{6} \tilde{\Phi}^2 + \frac{1}{6}
\left[ 4 \tilde{\Phi}_{,\mu}\tilde{\Phi}_{,\nu} -
\tilde{g}_{\mu\nu}
\tilde{\Phi}_{,\al} \tilde{\Phi}^{,\al} \right] + \frac{1}{3} \left[ \tilde{g}_{\mu\nu}
\tilde{\Phi} \Box \tilde{\Phi} -
\tilde{\Phi} \tilde{\Phi}_{;\mu\nu} \right] &=&  \tilde{T}_{\mu\nu}~,
\eea
which after contraction give a modified equation (\ref{eom_1})
\bea
\label{eom_1Ttil}
\left(\stackrel{\sim}{\Box} - \frac{1}{6} \tilde{R} \right) \tilde{\Phi} &=&
  \frac{\tilde{T}}{\tilde{\Phi}}~.
\eea
Note again that putting $\tilde{\Phi} = \tilde{\phi}_0 = \sqrt{6/8\pi G}$ into
(\ref{eom4Ttil}) gives the Einstein field equations, provided
$\tilde{T} = -(1/6) \tilde{R}\tilde{\Phi^2}$. However, as we will
see in a moment the {\it trace} of the energy-momentum tensor {\it must be
zero} in order to conserve the energy-momentum in a conformal frame.

Let us take the perfect fluid as a source of gravity with the
four-velocity $u^{\mu}$ ($u_{\mu}u^{\mu}=-1$), the energy density $\vr$
and the pressure ${\rm p}$, i.e.,
\bea
\label{enmom}
T^{\mu\nu} &=& (\vr+{\rm p})u^{\mu}u^{\nu}+{\rm p}g^{\mu\nu}~,
\eea
and transform it into a conformally related frame
\bea
\label{enmom_trafo}
\tilde{T}^{\mu\nu}
&=&
(\tilde{\vr}+\tilde{{\rm p}})\tilde{u}^{\mu}\tilde{u}^{\nu} +
\tilde{{\rm p}}\tilde{g}^{\mu\nu}~,
\eea
where
\bea
\tilde{u}^{\mu} &=& \frac{\d x^{\mu}}{\d \tilde{s}} =
\frac{1}{\Om}\frac{\d x^{\mu}}{\d s}=\Om^{-1}u^{\mu}~.
\eea
It is easy to note that the imposition of the conservation law in the first frame
\bea
\label{laws}
T^{\mu\nu}_{~~;\nu}=0~,
\eea
gives in the conformally related frame
\bea
\label{tildelaws}
 \tilde{T}^{\mu\nu}_{~~\tilde{;}\nu} &=& -\frac{\Om^{,\mu}}{\Om}\tilde{T}~.
\eea
From (\ref{tildelaws}) it appears transparent that the conformally
transformed energy-momentum tensor is conserved only if the trace of
it vanishes ($\tilde{T}=0$) \cite{jordan,weinberg}. For example,
in the case of barotropic fluid with
\bea
\label{barotropic}
{\rm p}&=&(\ga-1)\vr~ \hspace{0.5cm} \gamma = {\rm const.},
\eea
it vanishes only for radiation
${\rm p}=(1/3)\vr$~. This means that only the photons may obey
the equivalence principle and this is not the case for other types
of matter since with non-vanishing trace in (\ref{tildelaws}) we
deal with creation of matter process (compare with Self Creation Cosmology of Ref.
\cite{barber} which has the same field equations (\ref{eom4Ttil})
and (\ref{tildelaws}), but the equation (\ref{eom_1Ttil}) is the
same only for a vanishing curvature scalar $\tilde{R}$).

\section{Relation to Brans-Dicke and low-energy-effective superstring theories}
\label{relation}
\setcounter{equation}{0}

Due to the admission of a non-minimally coupled scalar field, conformal relativity
suggests its close relation to some other scalar-tensor theories of gravity such as
Brans-Dicke theory and the reduced low-energy-effective superstring theories.
It is widely known that Brans-Dicke theory (BD)~\cite{bd} was based on the ideas of
Jordan~\cite{jordan} and Mach~\cite{ray}. According to them, the inertial masses of
the elementary particles are not fundamental constants but the result
of the particles' interaction with the rest of the universe represented by
some cosmic field~\cite{weinberg}. In physical terms it can be formulated
by the fact that gravitational ``constant'' $G$ is related to
an average value of a scalar field $\Phi_{BD}$ which is not constant.
Due to some estimations concerning a simple form of a covariant field equation,
the size and mean mass density of the universe, the expectation value of the scalar
field $\Phi_{BD}$ is normalized, such as $\langle \Phi_{BD} \rangle \simeq
1/G$. These assumptions led Brans and Dicke to replace $G$ by
the inverse of the scalar field $1/\Phi_{BD}$ and to include an extra
energy-momentum tensor for the scalar field.

The resulting Brans-Dicke action reads as \cite{narlikar}
\bea
\label{BDaction}
S_{BD} &=& \frac{1}{16\pi} \int d^4x \sqrt{-g} \left[
\Phi_{BD} R - \frac{\om}{\Phi_{BD}}
{\partial}_{\mu}\Phi_{BD} {\partial}^{\mu}\Phi_{BD}
\right]~ + S_m,
\eea
where $\om$ is the Brans-Dicke parameter.

Varying the action (\ref{BDaction}) (with matter Lagrangian included) one
gets the field equations of the Brans-Dicke theory in the form
\bea
\label{BDequations}
\Box\Phi_{BD} &=& \frac{8\pi T}{3+2\om}~,\\
\label{BDequations1}
R_{\mu\nu}-\frac{1}{2}g_{\mu\nu}R &=& \frac{8\pi}{\Phi_{BD}}T_{\mu\nu}
+\frac{\om}{\Phi_{BD}^2}\left (\Phi_{BD,\mu}\Phi_{BD,\nu}
  -\frac{1}{2}g_{\mu\nu} \Phi_{BD,\rho}\Phi^{BD,\rho} \right )\nonumber \\
&& +\frac{1}{\Phi_{BD}} \left ( \Phi_{BD;\mu\nu} -g_{\mu\nu}\Box\Phi_{BD}\right
)~.
\eea
Independently, the conservation law for matter energy-momentum
tensor
\bea
T^{\mu\nu}_{;\nu} &=& 0~
\eea
may be imposed \cite{weinberg}.

In fact, the equation ({\ref{BDequations}) was obtained by
subtracting the equation of motion of the Brans-Dicke field
obtained from the action (\ref{BDaction})
\bea
2\Phi_{BD} \Box \Phi_{BD} - \Phi_{BD,\rho}\Phi^{BD,\rho} +
\frac{R}{\om} \Phi_{BD}^2 &=& 0~,
\eea
and the contracted equation (\ref{BDequations1}).
The Einstein limit of equations (\ref{BDequations})-(\ref{BDequations1})
is recovered for $\om~\rar~\infty$ \cite{weinberg}.

In order to relate conformal relativity with Brans-Dicke theory we
refer to conformally invariant actions (\ref{boundarytilde})
and (\ref{boundary}) in Jordan frame and define
\bea
\label{Phitophi}
\frac{1}{12} \Phi^2 &=& e^{-\phi}~,\\
\frac{1}{12} \tilde{\Phi}^2 &=& e^{-\tilde{\phi}}~,
\eea
which gives these actions in the form
\bea
\label{eff}
S &=& \int d^4x \sqrt{-g} e^{-\phi}\left[
\frac{1}{6} R + \frac{3}{2}
{\partial}_{\mu}\phi {\partial}^{\mu}{\phi}
\right]~,
\eea
\bea
\label{tildeeff}
\tilde{S} &=& \int d^4x \sqrt{-\tilde{g}} e^{-\tilde{\phi}} \left[
\frac{1}{6} \tilde{R}  + \frac{3}{2}
\stackrel{\sim}{\partial}_{\mu}\tilde{\phi}\stackrel{\sim}{\partial}^{\mu}\tilde{\phi}
\right]~.
\eea
These actions, however, are special cases of the Brans-Dicke
action which can be realized once one defines
\bea
\label{ephi}
\Phi_{BD} &=& e^{-\phi}
\eea
in (\ref{BDaction}) together with putting that $16\pi =1$, which
gives (\ref{BDaction}) in the form
\bea
\label{effom}
S &=&  \int d^4x \sqrt{-g} e^{-\phi}\left[
R  - \om
{\partial}_{\mu}\phi {\partial}^{\mu}{\phi}
\right]~,
\eea
so that one can immediately see that (\ref{eff}) and (\ref{effom})
are the same provided that the Brans-Dicke parameter
\bea
\label{32}
\om &=& - \frac{3}{2}~.
\eea
On the other hand, if one takes
\bea
\om &=& - 1
\eea
in (\ref{effom}), then one obtains the low-energy-effective
superstring action for only graviton and dilaton in the spectrum
\cite{lidsey,dab5,mar_string}
\bea
\label{effstr}
S &=& \int d^4x \sqrt{-g} e^{-\phi}\left[
R  + {\partial}_{\mu}\phi {\partial}^{\mu}{\phi}
\right]~.
\eea

In fact, the action (\ref{effom}) represents Brans-Dicke theory in
a special frame which is known as {\it string frame}. It is
because in superstring theory the coupling constant $g_s$ is related to
the vacuum expectation value of the dilaton by \cite{dab5,mar_string}
\bea
\label{gs}
g_s \propto e^{\phi/2}.
\eea

Now, the field equations which are obtain by the variation of
(\ref{effom}) with respect to the dilaton $\phi$ and the graviton
$g_{\mu\nu}$, respectively, are \cite{clw1}
\bea
\label{stringfe1}
R  + \om {\partial}_{\mu}\phi {\partial}^{\mu}{\phi} - 2\om \Box
\phi &=& 0~,\\
\label{stringfe2}
R_{\mu\nu} - \frac{1}{2} g_{\mu\nu} R &=& 8\pi e^{\phi} T_{\mu\nu}
+ (\om+1) {\partial}_{\mu}\phi {\partial}_{\nu}{\phi} -
\left( \frac{\om}{2} + 1 \right) g_{\mu\nu} {\partial}_{\rho}\phi {\partial}^{\rho}{\phi}
+ g_{\mu\nu} \Box \phi - \phi_{;\mu\nu}~.
\eea
Following the same track as for Brans-Dicke field equations now we
can contract (\ref{stringfe2}) and use (\ref{stringfe1}) to get
a similar equation to (\ref{BDequations}) in string frame, i.e.,
\bea
\label{strT}
{\partial}_{\rho}\phi {\partial}^{\rho}{\phi} - \Box \phi &=&
\frac{8\pi}{2\om +3} e^{\phi} T~.
\eea
Notice that putting $T=0$ in (\ref{strT}) (which is the case for dilaton-graviton
theory of superstring cosmology) we have that $ \Box \phi =
{\partial}_{\rho}\phi {\partial}^{\rho}{\phi}$ which gives exactly
the form of pre-big-bang field equations as presented in
\cite{clw1} with an arbitrary value of $\om$. On the other hand,
the limit $\omega \to - \frac{3}{2}$ of the equation (\ref{strT}) is
singular unless we take the trace of the energy-momentum tensor
$T=0$.

In fact, it can be just obtained by the application of (\ref{ephi}) into
(\ref{BDequations}) since
\bea
\Box \Phi_{BD} &=& e^{-\phi} \left( {\partial}_{\rho}\phi {\partial}^{\rho}{\phi}
- \Box \phi \right)~.
\eea
It is interesting to note that the Ricci scalar which can be
calculated from (\ref{stringfe1})-(\ref{stringfe2}) as
\bea
R_{\mu\nu} &=& 8\pi e^{\phi} T_{\mu\nu} - \phi_{;\mu\nu} + (\om + 1) \left(
{\partial}_{\mu}\phi {\partial}_{\nu}{\phi} -
g_{\mu\nu} {\partial}_{\rho}\phi {\partial}^{\rho}{\phi}
+ g_{\mu\nu} \Box \phi \right)~,
\eea
and for low-energy-effective superstring theory $\om = -1$ the whole lot of
its terms vanish. However, this is not the case in conformal
relativity $\om = -(3/2)$ for which this expression is not so simple and of
course from (\ref{strT}) one sees that the trace $T$ of the energy
momentum tensor must be zero in this limit.

In the context of superstring theory (with no matter fields) one often starts looking for
the exact solutions starting from the two equations \cite{clw2,bd97,veneziano}
\bea
\label{pbb1}
R  + \om {\partial}_{\mu}\phi {\partial}^{\mu}{\phi} - 2\om \Box
\phi &=& 0~,\\
\label{pbb2}
R_{\mu\nu} + \phi_{;\mu\nu} - (\om + 1) \left(
{\partial}_{\mu}\phi {\partial}_{\nu}{\phi} -
g_{\mu\nu} {\partial}_{\rho}\phi {\partial}^{\rho}{\phi}
+ g_{\mu\nu} \Box \phi \right) &=& 0~.
\eea
In the next section we will try to use these equations to present
cosmological solutions.

Now we come to an important remark. Namely, taking the limit $\om=-3/2$
in Brans-Dicke field equation (\ref{BDequations}) is singular
unless we assume that the trace of the energy-momentum tensor of
matter {\it vanishes}. The point is that, in fact, the conformal
relativity field equation (\ref{eom_1T}) is not an independent
equation from (\ref{eom4T}), as it is the case in Brans-Dicke
theory, but that it is exactly a contraction of (\ref{eom4T})!
Then, it may suggest that in order to get the proper limit of
Brans-Dicke theory from conformal relativity one should also
assume that in Brans-Dicke theory the trace of the energy-momentum
of matter should vanish. As we have mentioned, vanishing of the
energy-momentum for the perfect fluid requires its equation of
state for radiation.

Now let us move to the problem of frames. The different frames are
defined by the {\it coupling properties} of the scalar field
(dilaton) to gravity in the theory. In all three cases we have
discussed so far (conformal relativity, Brans-Dicke, superstrings)
we deal with {\it non-minimal} coupling of the scalar field to
gravity. For Brans-Dicke we call it {\it Jordan frame} and for
superstrings we call it {\it string frame}. For conformal
relativity we will also call it {\it Jordan frame} although we
deal with non-minimal coupling in both conformally related frames
(conformal invariance).

However, for all three theories one is usually interested in the
question of what is going on in the {\it Einstein frame} which is
defined as the frame in which the scalar field is {\it minimally}
coupled to gravity.

Let us now begin with the action (\ref{effom}) which admits an
arbitrary value of the parameter $\om$. It is easy to show \cite{clw2,bd97,lidsey}
that under a choice of a conformal factor
\bea
\label{Om_phi2}
\Omega &=& e^{-\frac{\phi}{2}}~,
\eea
the action (\ref{effom}) transforms into
\bea
\label{Eframe}
\tilde{S} &=& \int d^4x \sqrt{-\tilde{g}} \left[
 \tilde{R}  - \left( \om + \frac{3}{2} \right)
\stackrel{\sim}{\partial}_{\mu}\phi\stackrel{\sim}{\partial}^{\mu}\phi
\right]~
\eea
which for $\om=-3/2$ gives exactly the Einstein-Hilbert action
with no matter Lagrangian
\bea
\label{EHa}
\tilde{S} &=& \int d^4x \sqrt{-\tilde{g}} \tilde{R}~.
\eea
However, for low-energy-effective theory with $\om = -1$ this is
not Einstein-Hilbert action, but the Einstein gravity coupled
minimally to a non-vanishing scalar field $\phi$.
Notice that in terms of the $\Phi$ field defined in conformal
relativity according to (\ref{Phitophi}) the relation
(\ref{Om_phi2}) reads as
\bea
\label{Om_phi2Phi}
\Omega &=& e^{-\frac{\phi}{2}} = \frac{\Phi}{\sqrt{12}}~,
\eea
or, basically, $\tilde{\Phi}=$ const. in (\ref{PhitildetoPhi})
which has already been mentioned gives the Einstein limit of
conformal relativity. This means that the choice of the Einstein
frame is {\it unique} in conformal relativity and requires
$\tilde{\Phi}=$ const.

The appropriate field equations which come from (\ref{Eframe}) are
\bea
\label{Eframeeq1}
\tilde{R}_{\mu\nu}- \frac{1}{2}\tilde{g}_{\mu\nu}\tilde{R} &=&
\tilde{T}_{\mu\nu} \equiv \left( \om + \frac{3}{2} \right) \left[
2\stackrel{\sim}{\partial}_{\mu}\phi\stackrel{\sim}{\partial}_{\nu}\phi
- g_{\mu\nu} \stackrel{\sim}{\partial}_{\rho}\phi\stackrel{\sim}{\partial}^{\rho}\phi
\right]~,\\
\label{Eframeeq2}
\left( \om + \frac{3}{2} \right) \stackrel{\sim}{\Box} \phi &=&
0~.
\eea
From (\ref{Eframeeq1})-(\ref{Eframeeq2}) it follows that $\omega = -3/2$ is exactly the
border between a standard scalar field and a ghost (negative
kinetic energy) which is required in order to have an expansion minimum
in varying constant cosmologies \cite{barrowcyclic} and to mimic the phantom field
\cite{phantom03} which violates the weak energy condition $\tilde{\rho}+\tilde{p} <0$.
This is easily seen after calculating the energy-momentum tensor and
expressing the energy density and pressure in terms of scalar
field, i.e.,
\bea
\label{Eframerope}
\tilde{\rho} &=& \left( \omega + \frac{3}{2} \right)\dot{\phi}^2~,\\
\tilde{p} &=& \left( \omega + \frac{3}{2} \right)\dot{\phi}^2~.
\eea
This also means that, formally, the energy density and pressure vanish due to a
special choice of Brans-Dicke parameter $\omega$ and not
necessarily due to a vanishing of the scalar field in the Einstein
frame.

Obviously, a more general choice of the conformal factor which
leaves the action (\ref{effom}) invariant reads as an appropriate
generalization of the choice (\ref{PhitildetoPhi})
\bea
\label{Om_phidual}
\Omega &=& \frac{e^{-\frac{\phi}{2}}}{e^{-\frac{\tilde{\phi}}{2}}}
= \frac{\Phi}{\tilde{\Phi}}~,
\eea
and this is exactly the conformally invariant transformation
(\ref{PhitildetoPhi}) of the scalar field in conformal relativity.
Having applied (\ref{Om_phidual}) to (\ref{effom}) shows its
conformal invariance, i.e.,
\bea
\label{tildeff}
\tilde{S} &=&  \int d^4x \sqrt{-\tilde{g}} e^{-\tilde{\phi}}\left[
\tilde{R} + \frac{3}{2}
{\partial}_{\mu}\tilde{\phi} {\partial}^{\mu}{\tilde{\phi}}
\right]~.
\eea
In fact it works for any value of $\om$! Notice that conformal
invariance leads to a {\it self-duality transformation} \cite{lidsey}
known from superstring theory between the scalar fields
which leaves the action invariant and reads as
\bea
\tilde{\Phi} \to \Phi~,
\eea
or,
\bea
\tilde{\phi} \to \phi~,
\eea
or
\bea
\Omega \to \frac{1}{\Omega}~.
\eea

A slightly different way of getting conformal invariance of the action
(\ref{eff}) is the following conformal transformation
\cite{witten,lidsey}
\bea
\Omega &=& e^{-\phi}~,
\eea
which brings it to the form
\bea
\label{eff1}
S &=& \int d^4x \sqrt{-\tilde{g}} e^{\phi}\left[
\tilde{R} + \frac{3}{2}
{\partial}_{\mu}\phi {\partial}^{\mu}{\phi}
\right]~,
\eea
and it remains conformally invariant provided we replace
\bea
\phi \to - \tilde{\phi}~,
\eea
which gives
\bea
\label{eff2}
S &=& \int d^4x \sqrt{-\tilde{g}} e^{-\tilde{\phi}}\left[
\tilde{R} + \frac{3}{2}
{\partial}_{\mu}\tilde{\phi} {\partial}^{\mu}{\tilde{\phi}}
\right]~,
\eea
and this is another example of duality symmetry known from
superstring theory and may relate weak coupling regime with a
strong coupling regime of various superstring actions
\cite{witten}.

\section{Conformal cosmology in Einstein and string frames}
\label{CC}
\setcounter{equation}{0}

We discuss Friedmann cosmology in the two conformally related frames
as given in (\ref{conf_trafo}), i.e.,
\bea
\label{FRWtildemetric}
d\tilde{s}^2 &=&  - d\tit^2 + \tia^2 \left(\frac{dr^2}{1 - kr^2} + r^2
d\theta^2 + \sin^2{\theta} d\phi^2 \right)~\\
\label{FRWmetric}
ds^2 &=& - dt^2 + a^2 \left(\frac{dr^2}{1 - kr^2} + r^2
d\theta^2 + \sin^2{\theta} d\phi^2\right)~.
\eea
From (\ref{conf_trafo}), (\ref{FRWtildemetric}) and
(\ref{FRWmetric}) one can easily see that the time coordinates and
scale factors are related by \cite{clw1,bd97,polarski}
\bea
\label{cf1}
d\tit &=& \Omega dt~,\\
\label{cf2}
\tia &=& \Omega a~,
\eea
where for the full conformal invariance one has to apply the
definition of conformal factor (\ref{PhitildetoPhi}). However, in
the case of studying the Einstein limit we must take one of the
scalar fields constant, so that the conformal factor has the form
given by (\ref{Om_phi2}). This requires a replacement of (\ref{cf1})-(\ref{cf2})
into
\bea
\label{cf11}
d\tit &=& e^{-\frac{\phi}{2}} dt~,\\
\label{cf21}
\tia &=& e^{-\frac{\phi}{2}} a~,
\eea
where the quantities with tildes are in the Einstein frame while
those without tildes are in the string frame. In fact, the
transformations (\ref{cf1}) and (\ref{cf11}) are coordinate
transformations rather than conformal transformations and so they
are responsible for a difference in the time mesurements/scales in both
frames. Historically these scales were called atomic and cosmological,
for example \cite{dirac,narlikar}.

Imposing Friedmann metric (\ref{FRWmetric}) with in the Einstein
frame the equations (\ref{Eframeeq1})-(\ref{Eframeeq2}) give
\bea
\label{Efr1}
\left( \om + \frac{3}{2} \right) \left[\ddot{\phi} + 3 \frac{\tid}{\tia}\dot{\phi} \right] &=& 0~,\\
\label{Efr2}
3 \frac{{\tid}^2 + k}{\tia^2} &=&  \left( \om + \frac{3}{2} \right)
\dot{\phi}^2~,\\
\label{Efr3}
- 2 \frac{\tidd}{\tia} - \frac{{\tid}^2 + k}{\tia^2} &=&
\left( \om + \frac{3}{2} \right) \dot{\phi}^2~.
\eea
The solutions of (\ref{Efr1})-(\ref{Efr3}) for an arbitrary value
of the parameter $\om \neq - 3/2$ and $k=0$ read as
\bea
\tia &=& \mid \tit \mid^{\frac{1}{3}}~,\\
\phi &=& \phi_0 + \frac{1}{\sqrt{3(\om+\frac{3}{2})}} \ln{\mid
\tit \mid}~.
\eea
On the other hand, it is clear from (\ref{Efr1})-(\ref{Efr3}) that
for $\om = -3/2$ and $k=0$ the unique solution gives
\bea
\label{tid0}
\tid &=& 0~,
\eea
which is just a flat Minkowski universe. If we assume $k \neq 0$,
then we get from (\ref{Efr2}) that
\bea
\tia &=& \sqrt{-k} \hspace{5pt} \tit + \tit_0~,
\eea
which is admissible only for $k=-1$ and this solution represents
Milne universe \cite{narlikar} (in which there is no acceleration
of the expansion since $\ddot{\tilde{a}}=0$). However, its
relation to Minkowski spacetime requires coordinate
transformations which involves the two time scales - a dynamical
one and an atomic one \cite{narlikar} which may be responsible for
cosmological redshift effect. On the other hand, the solution for $k=+1$
would be possible only if the cosmological constant was admitted -
again, cosmological redshift in this Static Einstein model would
be the result of a different time scaling \cite{narlikar}.

In the string frame we use the Friedmann metric (\ref{FRWtildemetric})
which imposed into the equations (\ref{pbb1})-(\ref{pbb2}) for an arbitrary value of the
parameter $\om$ gives the following set of equations
\bea
\label{pbbom1}
\dot{\phi} - 3 \frac{\dot{a}}{a} &=& \frac{\ddot{\phi}}{\dot{\phi}}~,\\
\label{pbbom2}
- 3 \frac{\dot{a}^2 + k}{a^2} &=& - \left( \frac{\om}{2} + 1
\right) \dot{\phi}^2 + \ddot{\phi}~,\\
\label{pbbom3}
- 2 \frac{\ddot{a}}{a} - \frac{\dot{a}^2 + k}{a^2} &=&
\frac{\om}{2} \dot{\phi}^2 + \frac{\dot{a}}{a} \dot{\phi}~.
\eea
These equations (\ref{pbbom1})-(\ref{pbbom3}) give the following
solutions
\bea
\label{oms1}
a(t) &=& \mid t \mid^{\frac{3(\om+1) \pm
\sqrt{3(2\om+3)}}{3(3\om+4)}}~,\\
\label{oms2}
\phi(t) &=& \frac{-1 \pm \sqrt{3(2\om+3)}}{3\om+4} \ln{\mid t
\mid}~,
\eea
where following pre-big-bang/ekpyrotic scenario the solutions for
negative times are also admitted. From (\ref{pbbs1})-(\ref{pbbs2})
one can first find the pre-big-bang solutions for $\om=-1$
\cite{dab5} which are very well-known and read
\bea
\label{pbbs1}
a(t) &=& \mid t \mid^{\pm \frac{1}{\sqrt{3}}}~,\\
\label{pbbs2}
\phi(t) &=& (\pm \sqrt{3} - 1) \ln{\mid t \mid}~.
\eea
However, the conformal relativistic solutions for $\om = -\frac{3}{2}$
are
\bea
\label{cgrs1}
a(t) &=& \mid t \mid~,\\
\label{cgrs2}
\phi(t) &=& 2 \ln{\mid t \mid}~,
\eea
and show that they do not allow for two branches $`+'$ and $`-'$
and so they do not seem to allow scale factor duality \cite{meissven}
\bea
a(t) &\to& \frac{1}{a(-t)}, \hspace{0.4cm} \phi \to \phi - 6 \ln{a}~,
\eea
which is a cosmological consequence of string duality symmetries \cite{polchinsky}.
However, unlike pre-big-bang solutions (\ref{pbbs1})-(\ref{pbbs2})
which must be regularized at Big-Bang singularity because both the curvature and the
string coupling (\ref{gs}) diverge there, the solutions (\ref{cgrs1})-(\ref{cgrs2})
do not lead to strong coupling singularity in the sense of string
theory, since
\bea
g_s &=& e^{\frac{\phi}{2}} = \mid t \mid~,
\eea
and is regular for $t=0$. This has an interesting analogy with the
ekpyrotic/cyclic universe scenario where, in fact, the transition through
Big-Bang singularity takes place in the weak coupling regime
\cite{cyclic}. Note that form eqs. (\ref{cf11})-(\ref{cf21}) we
have
\bea
\tit &=& \tit_0 + \ln{t}~,
\eea
which according to (\ref{cgrs1}) gives
\bea
\tia &=& 1~.
\eea
This is consistent with what we have obtained in (\ref{tid0}) and
reflects the fact that in the Einstein frame the limit of an
expanding flat Friedmann metric is Minkowski universe.

\section{Discussion}
\label{disc}

We have shown that there is a simple relation between conformal
relativity (Hoyle-Narlikar theory) and a reduced (graviton-dilaton) low-energy-effective
superstring theory to Brans-Dicke theory. This relation shows that
the former is recovered from Brans-Dicke theory if one takes $\om = -3/2$
while the latter if one takes $\om =-1$. This may allow to study
the exact cosmological solutions of conformal relativity and its
properties appealing to the well-studied properties of the
Brans-Dicke theory and, in particular, to low-energy-effective superstring
effective theory.

In fact, Brans-Dicke parameter $\omega = -3/2$ gives a border between
a standard scalar field and a ghost (negative
kinetic energy) which is required in order to have an expansion minimum
in varying constant cosmologies \cite{barrowcyclic} and to mimic the phantom field
\cite{phantom03} which violates the weak energy condition $\rho+p <0$.

The conformally transformed energy-momentum tensor is conserved in conformal
relativity only if its trace vanishes. In the case of barotropic fluid
it vanishes only for radiation. This means that only the photons may obey
the equivalence principle and this is not the case for other types
of matter, since with non-vanishing trace we deal with creation of matter process.
We have compared that conformal relativity has lots of analogy with Self Creation Cosmology of Ref.
\cite{barber} which has the same field equations provided
we deal with cosmological models of vanishing curvature scalar.

We have presented basic cosmological solutions of conformal
relativity in both Einstein (minimally coupled) and string (nonminimally
coupled) frames. The Einstein limit
of flat conformal cosmological solutions is
unique and it is flat Minkowski space which requires the scalar field/mass
evolution instead of the scale factor evolution to explain
cosmological redshift \cite{annalen}. An interesting observation
is that like in ekpyrotic/cyclic models, a possible transition through
a singularity in conformal cosmology in the string frame takes place in
the weak coupling regime.

Some other interesting points can be made about the vanishing of some of
the PPN (parametrized-post-Newtonian) parameters \cite{mtw} in the theories
under consideration. It appears that
in conformal relativity only one of them ($\beta_1$) vanishes
while for superstring-effective theory two of them vanish ($\gamma$
and $\beta_4$).

Despite its restricted generality, conformally invariant theory of
gravity still seems to be attractive, for instance with respect to the
problem of a unique choice of vacuum in quantum field theory in curved
spaces \cite{birell}. There is no doubt that conformally
invariant theories still attract some attention in the context of
superstring theory and M-theory (see e.g. \cite{witten,guendelman,soda})
and require further studies.

\section{Acknowledgments}

MPD acknowledges a support from the Polish Research Committee grant No 2PO3B 090 23.


\begin{thebibliography}{99}

\bibitem{birell} Birell N.D. and Davies P.C.W. {\it Quantum field theory in
curved space}, Cambridge University Press, 1982.

\bibitem{HN} Hoyle, F., and Narlikar, J.V., Proc. Roy. Soc. A282
  (1964), 191~; {\it ibid} A294 (1966), 138 ~; {\it ibid} A270
  (1962), 334~.

\bibitem{chern} Chernikov, N., and Tagirov, E., Ann. Inst. Henri
  Poincar\`e 9 (1968), 109~.

\bibitem{be74} Bekenstein, J.D.,  Ann. Phys. (NY) 82 (1974) 535~.

\bibitem{narlikar} Narlikar, J.V., {\it Introduction to Cosmology},
  Jones and Bartlett Publishers, Inc. Portola Valley (1983)~.

\bibitem{penrose} Penrose, R., {\it Relativity, Groups and Topology},
  Gordon and Breach London (1964).

\bibitem{bbpp} Behnke, D., Blaschke, D.B., Pervushin, V.N., Proskurin,
  D.V., Phys. Lett. B {530} (2002), 20~.

\bibitem{iap} Blaschke, D., et al., in: {\it On the nature of dark
    energy}, eds. P. Brax, J. Martin, and J.-P. Uzan, IAP Paris
  (2002)~.

\bibitem{annalen} M.P. D\c{a}browski, D. Behnke and D. Blaschke,
in preparation.


\bibitem{weinberg_kosmkonst} Weinberg, S., Rev. Mod. Phys. 61 (1989), 23~.

\bibitem{apj517} Perlmutter, S., et al., ApJ 517 (1999), 565~.
\bibitem{aj116} Riess, A.G., et al., AJ 116 (1998), 1009 ~.
\bibitem{apj560} Riess, A.G., et al., ApJ 560 (2001), 49~.

\bibitem{ptw} Perlmutter, S., Turner, M.S., and White, M.,
  Phys. Rev. Lett. 83 (1999), 670~.

\bibitem{barber} Barber G.A., Gen. Rel. Grav. 14 (1982), 117;
Astroph. Space Sci. 282 (2002), 683.

\bibitem{pioneer} Anderson et al., Phys. Rev. D65 (2002), 082004.

\bibitem{bd} Brans, C., and Dicke, R.H., Phys. Rev. 124 (1961), 925~.

\bibitem{polchinsky} Polchinski J., {\em String Theory}, Cambridge
        University Press (1998).

\bibitem{dab5} D\c{a}browski, M.P., Ann. Phys. (Leipzig) 10
(2001), 195~.
\bibitem{mar_string} D\c{a}browski, M.P., {\it String Cosmologies}, University of
  Szczecin Press (2002)~.

\bibitem{dirac} Dirac P.A.M., Proc. Roy. Soc. A165 (1938), 199.

\bibitem{maeda} Y. Fujii and K.-I. Maeda, {\it The Scalar-Tensor
Theory of Gravitation}, Cambridge University Press (2003).

\bibitem{scaltens} Boisseau, B., Esposito-Farese, Polarski D., and
Starobinsky A.A., Phys. Rev. Lett. 85 (2000), 2236~.

\bibitem{polarski} Esposito-Farese, G., and Polarski, D., Phys.Rev.D
  63 (2001), 063504~.

\bibitem{mtw} Misner, C.W., Thorne, K.S., and Wheeler, J.A., {\it
    Gravitation}, W.H. Freeman and Company New York (1995)~.

\bibitem{hawk_ellis} Hawking, S.W., Ellis, G.F.R., {\it The
    large-scale structure of space-time}, Cambridge Univ. Press
  (1999)~.

\bibitem{flanagan} Flanagan \'E.\'E. gr-qc/0403063.


\bibitem{jordan} Jordan, P., Zeit. Phys. 157 (1959), 112~.

\bibitem{ray} d'Inverno, R., {\it Einf\"uhrung in die
    Relativit\"atstheorie}, VCH Weinheim (1995)~.

\bibitem{weinberg} Weinberg, S. {\it Gravitation and Cosmology}, John
  Wiley \& Sons New York (1972)~.

\bibitem{canuto} Canuto, V.M., Adams P.J., Hsieh S.H., and Tsiang
E., Phys. Rev. D16 (1977), 1643.


\bibitem{clw1} Copeland E.J., Lahiri A., and Wands D., Phys. Rev.
D50 (1994), 4868.

\bibitem{clw2} Copeland E.J., Lahiri A., and Wands D., Phys. Rev.
D51 (1995), 1569.

\bibitem{bd97} Barrow J.D. and D\c{a}browski M.P., Phys. Rev. D55
(1997), 630.

\bibitem{veneziano} Gasperini M. and Veneziano G., Phys. Rept.
(2002) .

\bibitem{lidsey} Lidsey, J., Wands D.W., Copeland E., Phys. Rept. 337 (2000), 343~.

\bibitem{barrowcyclic} Barrow J.D., Kimberley D., and Magueijo J.,
astro-ph/0406369.

\bibitem{phantom03} D\c{a}browski M.P., Stachowiak T. and Szyd{\l }owski M.,
      Phys. Rev. D{\bf 68} (2003), 103519.

\bibitem{witten} Witten, E., Nucl. Phys. B443 (1995), 85~.

\bibitem{meissven} Meissner K.A. and Veneziano G., Phys. Lett.
B267 (1991), 33; Mod. Phys. Lett. A6 (1991), 3397.

\bibitem{cyclic} Khoury J., Steinhardt P.J., Turok N., Phys. Rev.
Lett. 91 (2003), 161301; {\it ibidem} 92 (2004), 031302;
Khoury J., Ovrut B.A., Steinhardt P.J., Turok N., Phys. Rev. D64
(2001), 123522; {\it ibidem} D66 (2002), 046005.

\bibitem{guendelman} Guendelman E.I. and Spallucci E., {\it
Conformally invariant brane universe and the cosmological
constant}, e-print gr-qc/0402100.

\bibitem{soda} Kanno S., Sasaki M., and Soda J., Phys. Rev. D66
(2002), 683; Phys. Lett. B588 (2004), 203.

\end{thebibliography}
\end{document}